\documentclass[aoas,MSNbibl,nameyear,noautosecdot,rotating,dvips]{arximspdf}
\usepackage{dcolumn}
\usepackage{graphicx}
\doi{10.1214/11-AOAS481}
\volume{5}
\issue{4}
\pubyear{2011}
\firstpage{2448}
\lastpage{2469}

\makeatletter

\newcolumntype{d}[1]{D{.}{.}{#1}}

\makeatother

\begin{document}
\begin{frontmatter}

\title{A method for exploratory repeated-measures analysis applied to a
breast-cancer screening study}
\runtitle{Exploratory repeated-measures analysis}

\begin{aug}
\author[A]{\fnms{Adam R.} \snm{Brentnall}\corref{}\ead
[label=e1]{a.brentnall@qmul.ac.uk}},
\author[A]{\fnms{Stephen W.} \snm{Duffy}\ead
[label=e2]{s.w.duffy@qmul.ac.uk}},
\author[A]{\fnms{Martin J.} \snm{Crowder}\ead
[label=e3]{m.crowder@imperial.ac.uk}},
\author[A]{\fnms{Maureen G. C.} \snm{Gillan}\ead
[label=e4]{m.g.gillan@abdn.ac.uk}},
\author[A]{\fnms{Susan M.}
\snm{Astley}\ead[label=e5]{sue.astley@manchester.ac.uk}},
\author[A]{\fnms{Matthew G.}
\snm{Wallis}\ead[label=e6]{matthew.wallis@addenbrookes.nhs.uk}},
\author[A]{\fnms{Jonathan}
\snm{James}\ead[label=e7]{Jonathan.James@nuh.nhs.uk}},
\author[A]{\fnms{Caroline R. M.}
\snm{Boggis}\ead[label=e8]{caroline.boggis@nhs.net}} and
\author[A]{\fnms{Fiona J.} \snm{Gilbert}\ead
[label=e9]{f.j.gilbert@abdn.ac.uk}}
\runauthor{A. R. Brentnall et al.}
\affiliation{Queen Mary University of London,
Queen Mary University of London, Imperial College London,
University of Aberdeen, University of Manchester,
Addenbrookes Hospital, Cambridge, Nottingham City Hospital,
Wythenshawe Hospital, Manchester, and University of Aberdeen}
\address[A]{
Centre for Cancer Prevention\\
Wolfson Institute of Preventive Medicine\\
Barts and The London, Charterhouse square\\
London, EC1M 6BQ\\
United Kingdom\\
\printead{e1}} 
\end{aug}

\received{\smonth{6} \syear{2010}}
\revised{\smonth{2} \syear{2011}}

%
\begin{abstract}
When a model may be fitted separately to each individual statistical
unit, inspection of the point estimates may help the statistician to
understand between-individual variability and to identify possible
relationships. However, some information will be lost in such an
approach because estimation uncertainty is disregarded. We present a
comparative method for exploratory repeated-measures analysis to
complement the point estimates that was motivated by and is
demonstrated by analysis of data from the CADET II breast-cancer
screening study. The approach helped to flag up some unusual reader
behavior, to assess differences in performance, and to identify
potential random-effects models for further analysis.
\end{abstract}

%

\end{frontmatter}

\section{\texorpdfstring{Introduction.}{Introduction}}

In this article we propose an approach for exploratory
repeated-measures analysis. The term repeated measures is used in a
loose sense to mean that more than one datum is recorded on each
individual unit. However, the measurements themselves will be permitted
to have any data structure with a likelihood function, perhaps ranging
from replicated readings of the same quantity to multivariate
measurements of a stochastic process through time. The exploratory
method was motivated and is applied to data from the computer aided
detection evaluation trial (CADET) II trial, where 27 human readers
inspected distinct mammograms (breast $x$-rays) for cancer screening. Our
analysis aim is to determine whether real differences in behavior exist
between the individual readers, including whether any might be
outliers, and then if heterogeneity is observed, to seek possible
groups of similar individuals, and factors that correlate with the
differences. The proposal is partly motivated by the difficulty of such
an objective when the sample size is $27$, even when up to several
thousand measurements are observed on each reader.
The approach is developed in the next section and then it is
demonstrated using the data. Conclusions follow a section discussing
the application of the method to other data sets.

\section{\texorpdfstring{Method.}{Method}}
\label{secmethod}

The general data structure is first described and the main
simi\-larity-matrix idea is defined. Then, some properties of the matrix
are recorded and we comment on some ways in which it may be used for
exploratory repeated-measures analysis.

\subsection{\texorpdfstring{Setup.}{Setup}}

Suppose there are $n$ individual units ($i=1,\ldots,n$) with $n_i$
repeated measurements $\mathbf{y}_i = (y_{i,1},\ldots,y_{i,n_i})$
observed. The application in this paper has the units as humans who
interpret mammograms for cancer screening. We assume that there is a
suitable model form for the probability mass or density function
$p(y|\mathbf{u}_i)$ parametrized by $\mathbf{u}_i$ =
$(u_{i1},\ldots,u_{im})$, where~$m$ is the dimension of each
$\mathbf{u}_i$. For example, if the results are binary indicators
for recall ($y=1$) or no action ($y=0$), then $p(y|u_i)$ might be a~binomial model ($m=1$) with parameter $u_i$ interpreted as $i$'s
probability of recall. More generally, $p(y|\mathbf{u}_i)$ could be
developed from a data analysis, or knowledge of the problem, but we
assume that the $\mathbf{u}_i$ occur in the same form for each
individual $p(\mathbf{y}_i|\mathbf{u}_i)$. The statistical
modeling goal taken here is to understand variability of the
$\mathbf{u}_i$'s, perhaps through a model for~$p(\mathbf{u})$,
or~$p(\mathbf{u}|\mathbf{x})$ with explanatory variables
$\mathbf{x}_i = (x_{i1},\ldots,x_{ir})$. This two-stage model
structure taken is also taken in other areas, such as in applications
using linear mixed models [\citet{1990CrowderHand}].

Note that the setup considered is different to generalized estimating
equations (GEE). These are used to estimate marginal
(population-averaged) regression coefficients $\bolds{\beta}$ in a
repeated measures context where $\mathrm{E}(\mathbf{y}_i) = \mu
(\mathbf{x}_i;\bolds{\beta})$, but without assuming a full
probability model for $\mathbf{y}_i$ or even a~``true'' covariance
structure for~$\mathbf{y}_i$. In this paper we have a~full
(conditional) probability model for $\mathbf{y}_i$, $p(\mathbf
{y}_i | \mathbf{u}_i)$, that is based on subject-specific
parameters (random effects) and the focus is upon their distribution
over the population.

\subsection{\texorpdfstring{The similarity matrix.}{The similarity matrix}}\label{secmthdsim}

The exploratory measure that we call a similarity matrix is obtained in
two steps:
\begin{longlist}[(1)]
\item[(1)] Compute consistent $\hat{\mathbf{u}}_i$ for $i=1,\ldots,n$,
such as maximum likelihood estimates of $\mathbf{u}_i$; then
\item[(2)] calculate the $z$-matrix with row $i=1,\ldots,n$ and column
$j=1,\ldots,n$ entries from
%
\begin{equation}\label{eqnzij}
z_{ij} = \frac{p(\mathbf{y}_i|\hat{\mathbf{u}}_j)}{\sum
_{k=1}^n p(\mathbf{y}_i|\hat{\mathbf{u}}_k)}.
\end{equation}
\end{longlist}
A likelihood function $p(\mathrm{data}|\theta)$ reveals the relative
plausibilities of different parameter $\theta$-values in the light of
the data. Here $z_{ij} \propto p(\mathbf{y}_i|\hat{\mathbf
{u}}_j)$ does likewise for the $\mathbf{u}_j$ ($j=1,\ldots,n$) in
light of the data $\mathbf{y}_i$. The $z_{ij}$ quantity thus
explores the similarity of the $\mathbf{u}_j$'s via their
estimates, by measuring how close individual $j$'s parameter fit is to
individual $i$'s data.

If $y_{il}$ is a sequence of $l=1,\ldots,n_i$ binary indicators as
above, and a binomial likelihood is assumed for $p(y_i|u_i)$, then
using the notation $y_{i+} = \sum_{l=1}^{n_i}y_{il}$, we have
%
\begin{equation}
z_{ij} = \frac{\hat{u}_j^{y_{i+}}(1-\hat{u}_j)^{(n_i -
y_{i+})}}{\sum_{k=1}^n \hat{u}_k^{y_{i+}}(1-\hat{u}_k)^{(n_i -
y_{i+})}},
\end{equation}
because $n_i\choose y_i$ cancels in the numerator and denominator.

\subsection{\texorpdfstring{Some properties of the matrix.}{Some properties of the
matrix}}

\begin{longlist}[(13)]
\item[(1)]$0 \leq z_{ij} \leq1$.
\item[(2)]$z_{ij} = O(1/n)$. The practical significance is that larger
matrices will have smaller $z_{ij}$ terms.
\item[(3)]$z_{i+} = \sum_{j=1}^n z_{ij} = 1$.
\item[(4)]$z_{ij} \leq z_{ii}$ for $j\neq i$ if maximum-likelihood
estimation is used [because $p(\mathbf{y}_i | {\mathbf{u}})
\leq p(\mathbf{y}_i | \hat{\mathbf{u}}_i)$ for all
$\mathbf{u}$].
\item[(5)] The matrix follows from Bayes' rule
\[
p_e(\mathbf{u}|\mathbf{y}_i) \propto p(\mathbf
{y}_i|\mathbf{u})p_e(\mathbf{u}),
\]
where $p_e(\mathbf{u})$ is a probability mass function that
approximates variation of the random effect $\mathbf{u}$ across
individuals $p(\mathbf{u})$ by assigning mass $1/n$ to each of the
points $(\hat{\mathbf{u}}_1,\ldots,\hat{\mathbf{u}}_n)$. The
$e$ subscript is used in the notation to make explicit the reference to
this \textit{empirical} distribution. That is, $z_{ij} =
P_e(\mathbf
{u}_i = \hat{\mathbf{u}}_j | \mathbf{y}_i)$, and the $z_{ij}$
quantities are posterior $\mathbf{u}_i$ mass values where the
$\mathbf{u}$ distribution has been restricted to the points in
$p_e(\mathbf{u})$.
\item[(6)]$z$ is not symmetric unless $P_e(\mathbf{u}_j=\hat
{\mathbf
{u}}_i|{\mathbf{y}}_j)$ equals $P_e(\mathbf{u}_i=\hat
{\mathbf{u}}_j|{\mathbf{y}}_i)$. Thus, it is not a similarity
matrix in the usual sense.
\item[(7)] When $z_{ij} = z_{ii}$, then $\mathbf{y}_i$ is equally well
conditioned on $\hat{\mathbf{u}}_i$ and $\hat{\mathbf{u}}_j$,
and $P_e(\mathbf{u}_i = \hat{\mathbf{u}}_j| \mathbf
{y}_i) =
P_e(\mathbf{u}_i = \hat{\mathbf{u}}_i | \mathbf{y}_i)$.
\item[(8)] $z_{+j} > 1$ means that $\hat{\mathbf{u}}_j$ is very likely
the value for many $i$ and/or $z_{jj}$ is relatively large.
\item[(9)] An alternative measure $z_{+j}/z_{jj}$ can be used to assess the
importance of $\hat{\mathbf{u}}_j$ over the $i \neq j$.
\item[(10)] A measure of the overall concentration of the estimates is
$\operatorname{trace}(z)/\allowbreak n \in(0,1)$. Since $z_{++}=n$, $
\operatorname{trace}(z)/n$ attains maximum value $1$ when $z_{ii}=1$ for all $i$.
\item[(11)]$(z_{11},z_{22},\ldots,z_{nn})$, or $\operatorname{diag}(z)$ provides a
comparative measure of\break \mbox{concentration} in the estimates. This is because
point estimates $\hat{\mathbf{u}}_i$ with relatively high (or close
to 1) $z_{ii}$ entries may be interpreted as good predictions
since
%
\begin{eqnarray} \label{eqnprede}
\mathrm{E}_e(\mathbf{u}_i|\mathbf{y}_i) & = & \sum_{j=1}^n
\hat
{\mathbf{u}}_j P_e(\mathbf{u}_i = \hat{\mathbf{u}}_j|
\mathbf{y}_i) \nonumber\\[-8pt]\\[-8pt]
& = & \sum_{j=1}^n z_{ij}\hat{\mathbf{u}}_j.\nonumber
\end{eqnarray}\looseness=0
So for $z_{ii}$ close to one (and therefore $z_{ij}$ close to 0 for $j
\neq i$), a prediction from (\ref{eqnprede}) is likely to be very
close to $\hat{\mathbf{u}}_i$; for $z_{ii}$ not close to 1, the
point-estimate $\hat{\mathbf{u}}_i$ may be misleading because a
prediction from (\ref{eqnprede}) is subject to nonnegligible
averaging (shrinkage).
\item[(12)] If the $\mathbf{u}_i$ are distinct, then as $n_i \rightarrow
\infty$ for each $i$ the $z$-matrix will converge to the identity
matrix because a consistent estimator of $\mathbf{u}$ is used. In
practice, this means that when the $\mathbf{u}_i$ are different,
then a data set with large $n_i$ and well-estimated $\mathbf{u}_i$
will have a $z$-matrix close to the identity matrix. Conversely, little
structure is likely to be seen when all the $n_i$ are small, but it may
still be worth applying the method to see if this is the case. The most
useful case is likely to be when some of the $n_i$ are
mode-\break rate.
\item[(13)] A referee suggested a possible connection with Rubin's propensity
score [\citet{1983ROSENBAUMRubin}]. In that setting individuals are
matched (one a~case, the other a control) by a propensity score
$e(\mathbf{x}_i) = P(c_i=1|\mathbf{x}_i)$, where $c$ is an
indicator of being a case. In our setup individuals are matched to each
other through $z_{ij} = P_e({\mathbf{u}}_{i}=\hat{\mathbf
{u}}_j|\mathbf{y}_i)$. The propensity score reduces the
dimension of multivariate matching on $\mathbf{x}_i$ to a
univariate measure; the $z$-matrix transforms the dimensionality of
matching individuals on $\mathbf{u}_i$ to a~two-dimensional matrix.
\end{longlist}

\subsection{\texorpdfstring{Why use the matrix for exploratory analysis\textup{?}}{Why use the matrix for exploratory
analysis?}}

The first step in the computation of the $z$-matrix is to obtain point
estimates $(\hat{\mathbf{u}}_1, \ldots, \hat{\mathbf{u}}_n)$.
These might be plotted in exploratory analysis to look for clusters,
outliers and other structural relationships or trends across
individuals in the data. For example, one can plot the parameter fits
$\hat{\mathbf{u}}_i$ against each other, and against other
covariates by using a matrix scatter plot. An example demonstrating the
use of this approach for exploratory analysis with hierarchical linear
models is \citet{2005BowersDrake}.
One issue with the plots is that uncertainty in the point estimates is
disregarded and so apparent trends may be less impressive than first
appears, or masked by sampling variation.\looseness=1

A first way that the above properties of the $z$-matrix can be used to
add to the information in the plots is by helping to quantify the
concentration of each individual's estimate $\hat{\mathbf{u}}_i$ by
inspection of $\operatorname{diag}(z)$. A second way is by making comparisons
between the estimates $\hat{\mathbf{u}}_i$ and $\hat{\mathbf
{u}}_j$ from two individuals~$i$ and $j$, through the $z_{ij}$ and
$z_{ji}$ terms. An example of where these properties are useful is when
the $z_{ij}$ entries are zero, except for those within an identifiable
cluster of $\mathbf{u}$-values from the plots. This would suggest
that the individuals form a fairly homogeneous group. A third way is to
improve the point-estimates $\hat{\mathbf{u}}_j$ ($j=1,\ldots,n$)
themselves, by using equation (\ref{eqnprede}) to shrink the estimates
through $\mathrm{E}_e(\mathbf{u}_i|\mathbf{y}_i)$. A fourth way
is by using quantities such as $z_{+j} - z_{jj}$ or $z_{jj}/z_{+j}$ for
$j=1,\ldots,n$ to show the more important $\hat{\mathbf{u}}_j$, or
to identify outliers.
Some techniques to draw attention to these and other features are next
described.

\vspace*{3pt}\section{\texorpdfstring{Exploratory analysis with the $z$-matrix.}{Exploratory analysis with the
$z$-matrix}}

In this section we propose a~number of ways to present the $z$-matrix.
They will be demonstrated using the breast-screening data later on.

\subsection{\texorpdfstring{Tabular presentation of the matrix.}{Tabular presentation of the matrix}}\label{seczmtxplots}

When printing out the matrix it is important to display it in such a
way that important aspects of the data are clearly visible. With this
in mind we next suggest a way to display the matrix in tabular form:

\begin{itemize}
\item Print out the transpose of $z$, not $z$. When making comparisons
between individuals the main interest is comparing $z_{ij}$ for
$j=1,\ldots,n$. The transpose of the $z$-matrix is better because, as
in tables, it is easier to compare down columns than across rows
[\citet
{2004PData}].
\item Multiply the matrix by a power of 10 (e.g., 1,000) and
do not display (multiplied) values less than 1. The point here is to
focus the eye's attention on the difference between large, small and
negligible proportions by using the number of digits displayed in a
number. For example, the number 1,000 is seen to be larger than 10
because it has twice the number of digits; it is more difficult to see
at first glance that 0.1000 is bigger than 0.0010 because they have the
same number of digits. The choice of multiplication factor should
depend on $n$ since $z_{ij} = O(1/n)$.
\item Experiment with the order of individuals. The order used might be
based on an examination of one $z$-matrix, to regroup similar
individuals, or it might be made using the covariate $\mathbf{x}_i$
data. A recommended first order is by one of the $\mathbf{u}$
components, or a function of interest using the~$\mathbf{u}$'s.
\end{itemize}

\subsection{\texorpdfstring{Graphical presentation of the matrix.}{Graphical presentation of the
matrix}}

An alternative to printing the matrix is to use a plot. Since $\sum
_{j=1}^n z_{ij} = 1$, a recommended display is a histogram variety,
where there is one bar for each cell in the matrix. Such a~chart can be
produced using a symbols plot, with rectangles of area proportional to
$z_{ij}$. It is arguably easier to compare the shape of histograms down
a page (one for each of the $n$ units), so it might be better to leave
the matrix untransposed in this instance. Use of the transpose for
printing and the untransposed matrix for plotting might also help the
statistician to see different features.

\subsection{\texorpdfstring{Graphs to assess the number of groups.}{Graphs to assess the number of
groups}}

For scalar $\hat{u}_i$ (i.e., each component of $\hat
{\mathbf{u}}_i$ if a vector) order the individuals by $\hat{u}_i$.
Then a plot of ($\hat{u}_i$, $i/n$) provides the estimated distribution
function of $u$, based on the a priori $p_e(u)$.
Such an approach uses information in the separate $u$'s, but due to the
equal weights, it might be improved by using the data to change the
weights from $1/n$.
The proposal is to use an estimate of the density of $u_j$ from
$n^{-1}z_{+j}$. If the $u_i$'s are all well estimated and different,
then the weights will not change much from $1/n$. If some are more
likely over the sample than others, then they will be up-weighted, and
others will be down-weighted.
A~related quantity is the distribution function $Z_k = n^{-1}\sum
_{j=1}^k z_{+j}$.
Plots of the $z$-matrix density and distribution function can be used
to help assess the number of groups in the data.

\subsection{\texorpdfstring{Shrinking parameter fits.}{Shrinking parameter
fits}}

A way to incorporate estimation uncertainty into any exploratory plots
involving $\mathbf{u}$ is to use equation (\ref{eqnprede}) to
shrink the estimates through $\mathrm{E}_e(\mathbf
{u}_i|\mathbf
{y}_i)$. In this way, outliers might be more reliably identified, as
well as possible patterns.

\subsection{\texorpdfstring{Smoothing covariates.}{Smoothing covariates}}\label{secexasmoothx}

The matching of individuals through the $z$-matrix may be used to show
the average covariates $\mathbf{x}$ for a given $\hat{\mathbf
{u}}_i$. This might aid inspection of possible correlations beyond
using the observed covariates $\mathbf{x}_i$ recorded for each
individual $i=1,\ldots,n$ in plots against (functions of) parameters
$\hat{\mathbf{u}}_i$. We next show how $\tilde{\mathbf
{x}}_i =
\sum_{k=1}^n \mathbf{x}_k z_{ik}/z_{+k}$ can be derived as the
expected $\mathbf{x}$ given $\hat{\mathbf{u}}_i$ from the
$z$-approach.

Suppose we have data $\mathbf{d}$, known to be one of the
$\mathbf{y}_i$, but not which one, and have prior $P(\mathbf
{d}=\mathbf{y}_i) = 1/n$ for $i=1,\ldots,n$. If we were interested
in the probability that the data $i=1,\ldots,n$ were generated by
parameter fit $\hat{\mathbf{u}}_i$, then we could use
$P_e(\mathbf{d} = \mathbf{y}_i|\mathbf{u}=\hat
{\mathbf
{u}}_j) = z_{ij}/z_{+j}$.
Now
%
\begin{equation}
p_e(\mathbf{x}|\mathbf{u}) = \sum_{k=1}^n p(\mathbf
{x}|\mathbf{d}=\mathbf{y}_k, \mathbf{u}) P_e(\mathbf
{d}=\mathbf{y}_k|\mathbf{u}).
\end{equation}
In the case where the $\mathbf{x}_i$ are distinct, we model
$P(\mathbf{x}=\mathbf{x}_i|\mathbf{d}=\mathbf{y}_k,
\mathbf{u})$ by an empirical distribution so that $P(\mathbf
{x}=\mathbf{x}_i|\mathbf{d}=\mathbf{y}_k, \mathbf
{u}) =
1$ if $i=k$ for\vadjust{\goodbreak} $k=1,\ldots,n$, and $0$ otherwise, then $P(\mathbf
{x}=x_k|\mathbf{u}) = P(\mathbf{d}=\mathbf
{y}_k|\mathbf
{u})$, leading to
%
\begin{eqnarray} \label{eqnexpectedx}
\mathrm{E}_e(\mathbf{x}|\mathbf{u}=\hat{\mathbf{u}}_i)
& =
& \sum_{k=1}^n \mathbf{x}_k P_e(\mathbf{d}=\mathbf
{y}_k|\mathbf{u}=\hat{\mathbf{u}}_i) \nonumber\\[-8pt]\\[-8pt]
& = & \sum_{k=1}^n \mathbf{x}_k z_{ik}\big/z_{+k}.\nonumber
\end{eqnarray}
In the case where the $\mathbf{x}_i$ are not distinct, one can
still use $\tilde{\mathbf{x}}_i$ as defined above. A~crude way to
think of the approach is that individuals are locally clustered
depending on their $\hat{\mathbf{u}}$'s, and the average covariate
at that cluster is obtained. Thus, given $\hat{\mathbf{u}}$, the
variation in the $\tilde{\mathbf{x}}$'s is much less than the
original $\mathbf{x}$'s. It is hoped that the process will smooth
out some sampling variation, making it easier to assess if there are
any real patterns of interest between $\mathbf{x}$ and
$\mathbf{u}$.

\subsection{\texorpdfstring{Graphical testing.}{Graphical testing}}

A last exploratory approach is to follow \citet{2004Gelman},
\citet
{2009Bujaetal} and others by comparing $z$-matrices or associated
plots against null model simulations.

\section{\texorpdfstring{Background to application.}{Background to
application}}

Two human readers are presently used in England to interpret mammograms
(breast $x$-rays) from the breast-cancer screening program. This regimen
is often called double reading, but we will call it dual reading to
emphasize that two independent readers inspect each mammogram. If both
readers find no abnormalities, the screenee is notified of the negative
result and no further action is taken. If both readers find a
suspicious abnormality, the screenee is recalled for further
investigations. If the readers disagree, one common practice is to have
a third reader arbitrate. Typically, for 1,000 women undergoing
screening, around 42 might be recalled, of whom 8 are found to have
cancer after further investigation [\citet{2009DoHScreening}]. Several
studies have shown that two readers can detect more cancers than a
single reader [\citet{2008TaylorPotts}]. The computer aided detection
evaluation trial (CADET) II was designed to assess whether a single
reader using a computer-aided detection tool could match the
performance of two readers.

In the trial 31,057 mammograms were read at three centers in England,
such that a ratio of $1:1:28$ were, respectively, dual reading only;
single-reading with CAD (computer-aided detection) only; and both dual
reading and single reading with CAD. Most of the screens were therefore
matched pairs from dual reading and single reading with CAD. The reason
why some screens were only read by one of the regimens was to reduce
the possibility of bias from readers changing their behavior due to the
knowledge that a~further\vadjust{\goodbreak} reading of the case would take place. Only the
28,204 matched-pair cases are considered from now on.
The main detection result was that 199 out of the 227 cancers detected
were recalled by dual reading, and 198 by single reading with CAD. 170
of the cases were detected by both, so the single readers with CAD
detected 28 cases missed by dual reading; dual reading detected 29
cases missed by the single reader with CAD (and $170+28+29=227$). The
overall recall rate for dual reading was 3.4\% and for single reading
with CAD it was slightly higher at 3.9\%. The analysis of the trial in
\citet{2008Gilberetal} found that single reading with computer-aided
detection could be an alternative to dual reading.

The primary analysis published in \citet{2008Gilberetal}
addressed the
question of whether detection and recall rates differ between dual
reading versus single reading with CAD. Further questions may be posed
of the data to help improve best practice in other areas: if factors
can be identified that predict outcomes prior to the screen being read,
then steps might be taken to mitigate risks. The aim of the analysis in
this article is to assess whether individual readers behaved
differently, and to determine if any factors might influence whether a
reader missed more cancers, or recalled more often than others. In the
data available from the trial we had information on their training
(radiologist, radiographer, other) and the number of years they had
read mammograms prior to the trial, and we explore whether any
differences between them might be related to these two factors.
Although there are a~large number of screens, the total number of
readers involved in the trial was $27$, and so drawing inference is
more difficult than might appear from consideration of the large number
of 28,204 cases.

\section{\texorpdfstring{Reader recall and detection rates.}{Reader recall and detection rates}}\label{secanal1}

In this section we use data from CADET II to demonstrate the $z$-matrix
exploratory analysis as a precursor to model building. The aim of the
analysis is to explore the data to assess if and why some readers
performed differently to others.

\subsection{\texorpdfstring{Data.}{Data}}

We present two exploratory analyses, one for the first reader in a
dual-reader pair, the second for a single reader with CAD. In the case
of a first reader $i$ from a dual reading the response is detection of
cancer: $y=1$ when a cancer is detected, 0 otherwise. In the case of a
single reader~$i$ with CAD the response is recall: $y=1$ if a case is
recalled, 0 if not. There are $k=1,\ldots,n_i$ screens by individual
$i$, and we take
\[
p(\mathbf{y}_i|u_i) \propto u_i^{y_{i+}}(1-u_i)^{n_i - y_{i+}},
\]
thus assuming that the $y_{ik}$ are conditionally independent with
$P(y_{ik}=1) = u_i$.
These data are shown in Tables \ref{tbldrdr1} and \ref{tblrecallcad}.
The total number of readers in both regimens differs partly due to not
all of them being trained to used CAD. Further details about the data
are in \citet{2008Gilberetal}.

\begin{table}
\caption{Number of cases detected and recalled by first 26 dual
readers. Analysis is undertaken for~detection~rate $u_i$}\label{tbldrdr1}
\begin{tabular*}{\tablewidth}{@{\extracolsep{\fill}}ld{3.0}d{3.0}rd{2.1}d{3.0}@{}}
\hline
& \multicolumn{1}{c}{\textbf{Cancers}}
&  & \multicolumn{1}{c}{\textbf{Screens}\hspace*{-2pt}}
& \multicolumn{1}{c}{\textbf{MLE (\%)}} & \multicolumn{1}{c@{}}{\textbf{Concentration}} \\
\multicolumn{1}{@{}l}{\textbf{Center}} & \multicolumn{1}{c}{$\bolds{y_{i+}}$} & \multicolumn{1}{c}{\textbf{Recalls}}
& \multicolumn{1}{c}{$\bolds{n_i}$\hspace*{-2pt}}
& \multicolumn{1}{c}{$\bolds{\hat{u}_i = y_{i+}/n_i}$}
& \multicolumn{1}{c@{}}{$\bolds{(z_{ii} \times
1\mbox{\textbf{,}}000)}$}\\
\hline
2 & 2 & 10 & 18 & 11.1 & 290 \\
2 & 8 & 26 & 92 & 8.7 & 375 \\
2 & 4 & 19 & 53 & 7.5 & 363 \\
3 & 5 & 11 & 355 & 1.4 & 108 \\
1 & 5 & 16 & 394 & 1.3 & 92 \\
3 & 9 & 27 & 805 & 1.1 & 103 \\
2 & 11 & 36 & 1,022 & 1.1 & 109 \\
1 & 15 & 62 & 1,412 & 1.1 & 124 \\
1 & 6 & 24 & 628 & 1.0 & 73 \\
2 & 18 & 76 & 1,922 & 0.9 & 124 \\
2 & 11 & 46 & 1,384 & 0.8 & 83 \\
2 & 14 & 67 & 2,128 & 0.7 & 82 \\
1 & 8 & 62 & 1,221 & 0.7 & 68 \\
3 & 5 & 25 & 769 & 0.7 & 60 \\
1 & 1 & 3 & 160 & 0.6 & 47 \\
3 & 6 & 29 & 997 & 0.6 & 64 \\
3 & 12 & 61 & 2,002 & 0.6 & 76 \\
2 & 7 & 34 & 1,180 & 0.6 & 66 \\
3 & 5 & 23 & 906 & 0.6 & 63 \\
3 & 7 & 27 & 1,312 & 0.5 & 69 \\
1 & 8 & 51 & 1,571 & 0.5 & 74 \\
1 & 10 & 57 & 2,132 & 0.5 & 86 \\
2 & 7 & 40 & 1,556 & 0.4 & 82 \\
1 & 3 & 21 & 735 & 0.4 & 72 \\
3 & 8 & 48 & 2,166 & 0.4 & 124 \\
1 & 4 & 46 & 1,284 & 0.3 & 127
\\[4pt]
OVERALL & 199 &947& 28,204 & 0.7 & \multicolumn{1}{c@{}}{$-$} \\
\hline
\end{tabular*}
\end{table}

\begin{table}
\caption{Number of cases detected and recalled by 18 computer-assisted
readers. Three cancer cases from center 2 had a missing reader
identifier. Analysis is carried out on recall rate $u_i$} \label{tblrecallcad}
%
\begin{tabular*}{\tablewidth}{@{\extracolsep{\fill}}ld{3.0}d{3.0}rcc@{}}
\hline
&
& \multicolumn{1}{c}{\textbf{Recalls}} & \multicolumn{1}{c}{\textbf{Screens}\hspace*{-2pt}}
& \multicolumn{1}{c}{\textbf{MLE (\%)}} & \multicolumn{1}{c@{}}{\textbf{Concentration}} \\
\multicolumn{1}{@{}l}{\textbf{Center}} & \multicolumn{1}{c}{\textbf{Cancers}} & \multicolumn{1}{c}{$\bolds{y_{i+}}$}
& \multicolumn{1}{c}{$\bolds{n_i}$\hspace*{-2pt}}
& \multicolumn{1}{c}{$\bolds{\hat{u}_i = y_{i+}/n_i}$}
& \multicolumn{1}{c@{}}{$\bolds{(z_{ii} \times
1\mbox{\textbf{,}}000)}$}\\
\hline
2 & 11 & 57 & 953 & 6.0 & 170 \\
2 & 11 & 64 & 1,080 & 5.9 & 170 \\
2 & 14 & 59 & 1,012 & 5.8 & 160 \\
2 & 7 & 61 & 1,062 & 5.7 & 156 \\
2 & 9 & 69 & 1,257 & 5.5 & 156 \\
2 & 9 & 49 & 921 & 5.3 & 143 \\
1 & 16 & 113 & 2,408 & 4.7 & 257 \\
2 & 8 & 46 & 993 & 4.6 & 168 \\
2 & 5 & 46 & 1,037 & 4.4 & 183 \\
1 & 12 & 87 & 2,150 & 4.0 & 322 \\
2 & 5 & 36 & 1,037 & 3.5 & 172 \\
1 & 11 & 76 & 2,266 & 3.4 & 249 \\
3 & 9 & 61 & 2,045 & 3.0 & 171 \\
1 & 17 & 79 & 2,713 & 2.9 & 180 \\
3 & 7 & 27 & 953 & 2.8 & 141 \\
3 & 25 & 84 & 3,089 & 2.7 & 188 \\
3 & 9 & 48 & 1,835 & 2.6 & 183 \\
3 & 10 & 35 & 1,390 & 2.5 & 192 \\
[4pt]
2 (Unk.) & 3 & 3 & 3 & $-$ & $-$ \\[4pt]
OVERALL & 198& \multicolumn{1}{r}{1,097\hphantom{00}} & 28,204 & 3.9 & $-$\\
\hline
\end{tabular*}
\end{table}

Exploratory analyses were also conducted on other combinations of
interest, such as on detection rate for single readers with CAD and for
second readers in a dual-reader pair. The two presented are chosen
because they show how the $z$-matrix can help to identify similar
groups of readers.

\subsection{\texorpdfstring{Exploratory analysis with similarity matrix.}{Exploratory analysis with similarity
matrix}}

\begin{figure}
\begin{tabular}{@{}c@{}}

\includegraphics{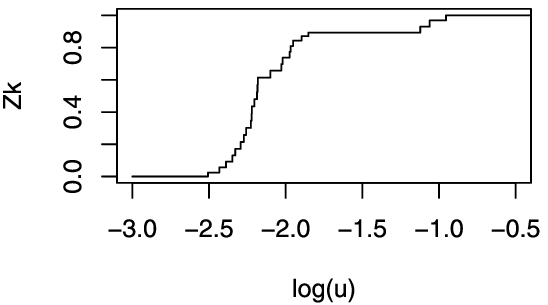}
\\
(a)\\[4pt]

\includegraphics{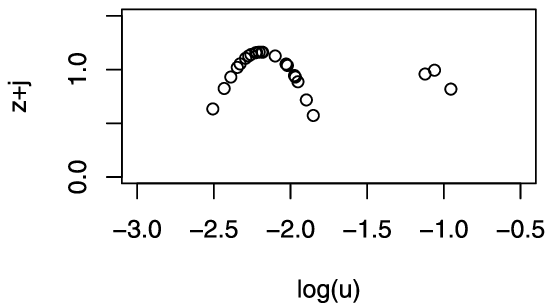}
\\
(b)
\end{tabular}
\caption{Exploratory plots for detection rate, first dual reader. On
the $x$-axis are $\hat{u}_i$ on a $\operatorname{log}_{10}$ scale. The
$y$-axis in \textup{(a)} is $Z_k$ as defined in Section \protect\ref
{seczmtxplots}, on
\textup{(b)} it is $z_{+j}$, which corresponds to the jump sizes in \textup{(a)}.}
\label{figxtraplot1}
\vspace*{-3pt}
\end{figure}

For detection rate the point estimates $\hat{u}_i$ in Table \ref
{tbldrdr1} show a group of three individuals with much higher
detection rates than the others. Although these readers saw a
relatively small number of cases, and the numbers detected are not
larger than other readers, the $z$-matrix [Supplementary Table 1,
Brentnall et al. (\citeyear{2011Brentnalletal})]
suggests the differences are not due to chance. The $z$-matrix has a
block structure with one block corresponding to the 3 outlying readers,
and the other block to everyone else. This can also be seen in Figure
\ref{figxtraplot1}, which contains plots introduced in Section \ref
{seczmtxplots}. The charts suggest that the results might be too
extreme to be due to random variation, and that there are two groups.
Further evidence of this is seen in the $z_{ii}$ measures of
concentration for first dual reader detection rate, shown in Table \ref
{tbldrdr1}: are all low.

For single-readers with CAD the $z$-matrix based on data in Table \ref
{tblrecallcad} is shown in Table \ref{tblcadZmtx}. The number of
digits reading down each column gives an impression of the size of each
$z_{ij}$ for $j=1,\ldots,18$, and the table shows a~center effect where
the clearest difference is between centers 2 and 3, with readers from
center 1 straddling the two.

\subsection{\texorpdfstring{Model.}{Model}}

The exploratory analysis suggests that a continuous model of $p(u)$
might be inappropriate because there appear to be clusters, or at least
there is not enough information to separate individuals within the
clusters. Therefore, a more plausible approach than a continuous
distribution is to take a discrete distribution for $u$, with unknown
locations $u_j$ and masses~$\theta_j$ for $j=1,\ldots,k$, where $k
\leq
n$. That is, $P(u=u_j;\bolds{\theta}) = \theta_j$. The
nonparametric maximum likelihood (NPML) estimate of $p(u)$ is a
discrete distribution and has the benefit of not requiring
specification of the form of~$p(u)$. An expectation-maximization (EM)
algorithm is used to next obtain the NPML estimates [\citet
{1978Laird}], and a likelihood-ratio test is used to compare the model
fit against a null model with a single atom. The $p$-values presented
follow \citet{1987SelfLiang}, and are used as a way to show the
evidence for the fitted model, rather than to formally control type I
error. This is relevant because the test is \textit{post hoc} based on
exploratory analysis, so there is an element of multiple testing.

%
\begin{sidewaystable}
\textwidth=\textheight
\tablewidth=\textwidth
\caption{($z^T \times1\mbox{,}000$)-matrix for reader recall rates using CAD.
Within center the individuals are ordered ascending by $\hat{u}_i$.
Note that the $z$-matrix is transposed in all the tables in this
article, and so, for example, the diagonal is the largest value down
each column, not row} \label{tblcadZmtx}
\begin{tabular*}{\tablewidth}{@{\extracolsep{\fill}}ld{3.0}d{3.0}d{3.0}
d{3.0}d{3.0}d{3.0}d{3.0}d{3.0}d{3.0}d{3.0}d{3.0}d{3.0}d{3.0}d{3.0}
d{3.0}d{3.0}d{3.0}c@{}}
\hline
\multicolumn{1}{@{}l}{\textbf{Center}}&\multicolumn{1}{c}{\hspace*{-2pt}\textbf{3}}
&\multicolumn{1}{c}{\hspace*{-2pt}\textbf{3}}
&\multicolumn{1}{c}{\hspace*{-2pt}\textbf{3}}&\multicolumn{1}{c}{\hspace*{-2pt}\textbf{3}}
&\multicolumn{1}{c}{\hspace*{-2pt}\textbf{3}}&\multicolumn{1}{c}{\hspace*{-2pt}\textbf{1}}
&\multicolumn{1}{c}{\hspace*{-2pt}\textbf{1}}&\multicolumn{1}{c}{\hspace*{-2pt}\textbf{1}}
&\multicolumn{1}{c}{\hspace*{-2pt}\textbf{1}}&\multicolumn{1}{c}{\hspace*{-2pt}\textbf{2}}
&\multicolumn{1}{c}{\hspace*{-2pt}\textbf{2}}&\multicolumn{1}{c}{\hspace*{-2pt}\textbf{2}}
&\multicolumn{1}{c}{\hspace*{-2pt}\textbf{2}}&\multicolumn{1}{c}{\hspace*{-2pt}\textbf{2}}&
\multicolumn{1}{c}{\hspace*{-2pt}\textbf{2}}&\multicolumn{1}{c}{\hspace*{-2pt}\textbf{2}}&
\multicolumn{1}{c}{\hspace*{-2pt}\textbf{2}}&\multicolumn{1}{c@{}}{\textbf{2}}\\
\hline
3&192&176&146&117&73&80&13&&&31&&&&&&&&\\
3&187&183&176&130&102&115&27&&&45&1&&&&&&&\\
3&172&176&188&138&132&150&50&1&&62&1&1&&&&&&\\
3&148&156&174&141&158&175&87&2&&84&3&1&&&&&&\\
3&111&117&128&136&171&176&149&7&&115&7&3&&&&&&\\
1&129&136&152&140&168&180&118&4&&100&5&2&&&&&&\\
1&37&35&24&93&109&77&249&72&1&169&33&18&1&&&&&\\
1&2&1&&19&7&1&57&322&75&109&150&111&25&7&5&4&2&\hphantom{00}3\\
1&&&&2&&&2&112&257&26&169&168&97&67&46&41&31&\hphantom{0}33\\
2&24&21&11&76&80&47&238&117&2&172&49&27&2&&&&&\\
2&&&&5&1&&8&216&214&51&183&161&64&34&22&19&13&\hphantom{0}15\\
2&&&&3&&&2&135&255&31&174&168&89&58&39&35&26&\hphantom{0}28\\
2&&&&&&&&7&97&3&78&104&143&151&130&124&116&114\\
2&&&&&&&&3&55&2&56&80&139&156&147&143&140&137\\
2&&&&&&&&1&19&1&31&50&122&145&156&159&164&162\\
2&&&&&&&&&12&&25&42&114&137&155&160&168&167\\
2&&&&&&&&&8&&19&34&104&126&152&159&170&170\\
2&&&&&&&&&6&&16&30&99&119&148&157&169&170\\
\hline
\end{tabular*}
\end{sidewaystable}
%

\subsection{\texorpdfstring{Results.}{Results}}

For the first reader, the EM-algorithm fit has just two atoms at
(0.0066, 0.0855) with respective masses (0.891, 0.109) and
log-likelihood $-1\mbox{,}170$.151. This compares against a null model with a
single point $0.0071$ and log-likelihood $-1\mbox{,}184$.125. A likelihood-ratio
test to compare the models rejects the hypothesis of no difference,
with $p\mbox{-value} <0.001$. The estimation results for $p(u)$ corroborate
the exploratory analysis: the first location (0.0071) is for the
majority of readers (the mass is 0.891); the second location (0.0855)
is for the top 3 readers in Table \ref{tbldrdr1} with much higher
detection rates.

The model fit for recall rate by readers using CAD also confirms the
exploratory analysis. There are two points at (0.0293, 0.0507) with
respective mass (0.449 0.551) and log-likelihood $-$4,606.186. The
degenerate fit is $0.0389$ and has log-likelihood $-$4,637.097, so a
likelihood-ratio $p\mbox{-value} <0.001$.\vadjust{\goodbreak}

\subsection{\texorpdfstring{Interpretation.}{Interpretation}}\label{secanal1int}

The unusual group of three readers' detection rates, within the same
center, can be explained by job title: they were the only radiographers
in that center. However, it is unlikely that radiographers are assigned
more cancer cases than radiologists because the outcome is unknown
prior to the screening. It seems more likely that center 2 used a~post-event method of deciding who to call the first reader. This is
discussed further in the next section.

Single readers with CAD were found overall to have higher recall rates
(3.9\%) than dual reading (3.4\%). Table \ref{tblrecallcad} shows that
most of the readers with higher recall rates were in center 2. The
$z$-matrix in Table \ref{tblcadZmtx} and the model estimation results
point toward a difference that is linked to center~2. That is, the
slight overall increase in recall rate of single reading with CAD over
dual reading might have been caused by a policy difference, or
difference in case-mix at one of the centers rather than errant
individual readers.

\section{\texorpdfstring{Categorical dual-reader outcomes.}{Categorical dual-reader outcomes}}\label{secanal2}

The analysis in the previous section focused on binary outcomes. One of
the advantages of the similarity matrix for exploratory analysis is
that it can be readily applied to any likelihood model $p(\mathbf
{y}|\mathbf{u})$. In this section we show an exploratory analysis
of dual-reader performance when 6 categorical outcomes are considered
and the likelihood of multinomial form.

\subsection{\texorpdfstring{Data.}{Data}}

In this analysis each screenee belongs to a state $S_{lm}$, where
$l=1,2$, respectively, denote a decision to recall or not by a reader
from the dual-reading regimen; $m=1$ for cancer present, $m=2$ for
cancer absent and $m=3$ for cancer unknown. Thus, $P(y_{ik}=S_{lm}) =
u_{ilm}$ for each case $k=1,\ldots,n_i$ seen by reader $i$.

The different states arise because even if a reader does not flag (or
does flag) a~case for recall, they may (or, respectively, may not) be
recalled in the trial. More specifically, when a case was flagged for
recall and it was recalled for further tests it is known whether there
was a cancer. When it was not recalled by the dual readers or the
single reader with CAD we call it ``unknown'' because no further tests
are undertaken, but the vast majority of such cases will not have a
cancer present. Cases that are flagged for recall but are unknown were
not recalled after arbitration, or were not flagged for recall by the
single reader with CAD. Cases that were not flagged for recall but the
outcome is known might have been recalled after arbitration, or
recalled by the single reader with CAD.

The data are found in Supplementary Table 2.
All dual readers are included, so we ignore whether the reader was
marked as a first or second reader. Some of the second reader
identifiers at center 2 were missing, but all data were available for
the other centers.

\subsection{\texorpdfstring{Exploratory analysis.}{Exploratory
analysis}}

Inspection of a scatter-matrix plot of $\hat{u}_{ijk}$ values for the
states $S_{lm}$ and reader experience, over readers $i=1,\ldots,27$,
does not show any clear trends (Supplementary Figure 1)
apart from a possible difference between the centers. However, a
pattern is present in $S_{13}$ vs. experience but it is masked
somewhat by between-center differences and sampling variation. The
following exploration of the $z$-matrix in conjunction with the data
helped to determine if there were any systematic differences between
the readers, and to identify and show more clearly the correlation
between experience and $S_{13}$.

\begin{table}
\caption{($z^T \times1\mbox{,}000$)-matrix for dual-reader categorical
outcomes within center 1. The data are in Supplementary Table 2}
\label{tblzijcat1}
%
\begin{tabular*}{\tablewidth}{@{\extracolsep{\fill}}ld{3.0}d{3.0}d{3.0}
d{3.0}d{3.0}d{3.0}d{3.0}@{\hspace*{-2pt}}}
\hline
\multicolumn{1}{@{}l}{\textbf{Experience: }
\hspace*{2.4pt}\textbf{5}}
&\multicolumn{1}{c}{\hspace*{-2pt}\textbf{14}}&\multicolumn{1}{c}{\hspace*{-2pt}\textbf{0.5}}
&\multicolumn{1}{c}{\hspace*{-2pt}\textbf{4}}&\multicolumn{1}{c}{\hspace*{-2pt}\textbf{6}}
&\multicolumn{1}{c}{\hspace*{-2pt}\textbf{12}}&\multicolumn{1}{c}{\hspace*{-2pt}\textbf{15}}
&\multicolumn{1}{c@{}}{\textbf{18}}\\
\hline
\multicolumn{1}{@{}l}{\hphantom{\textbf{Experience: }}993}&&3&&&&&\\
&997&13&&&&1&\\
\multicolumn{1}{@{}l}{\hphantom{\textbf{Experience: 00}}7}&3&973&&&1&4&2\\
&&8&579&153&156&356&20\\
&&&50&655&94&117&22\\
&&&29&59&666&56&39\\
&&3&341&133&82&465&24\\
&&&&&&&892\\
\hline
\end{tabular*}
\end{table}

\begin{figure}
\begin{tabular}{@{}cc@{}}

\includegraphics{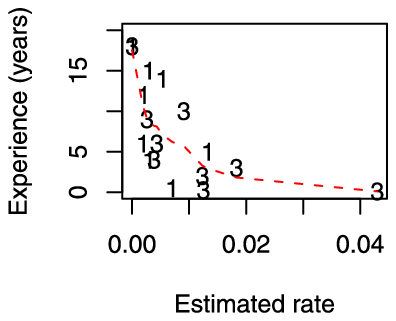}
 & \includegraphics{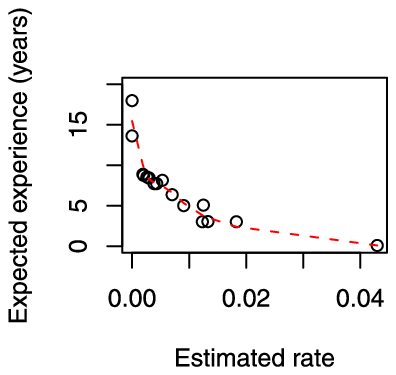} \\
(a) & (b)\\[4pt]

\includegraphics{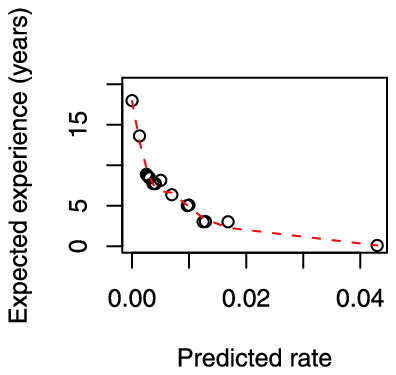}
 & \includegraphics{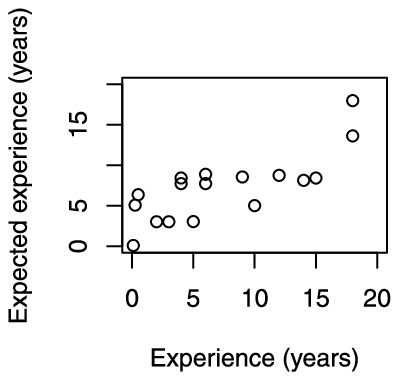} \\
(c) & (d)
\end{tabular}
\caption{Exploratory plots for recall-but-overruled rate vs.
reader experience (years), based on data from Supplementary Table 2
($S_{13}$ vs. the rest). The data show some evidence that
dual readers with less experience are more likely to be overruled in
centers 1 and 3. Readers from center 2 are excluded from the plots
because that center did not record when a reader flagged a case for
recall that was not recalled (all $S_{13}=0$). Plot \textup{(a)} shows
experience against the original estimates $\hat{u}$, with center number
as the symbol. Plot \textup{(b)} uses expected experience on the $y$-axis,
following the approach in Section \protect\ref{secexasmoothx}. The
plots are
presented with experience on the $y$-axis because they show a quantity
for the expected experience given $u$. Plot \textup{(c)} replaces the original
estimate $\hat{u}$ by a prediction from equation (\protect\ref{eqnprede}).
Each dashed line (-- --) is a loess smoother fit. Plot \textup{(d)} shows how
expected experience relates to the original data.} \label{figonon}
\end{figure}

\begin{figure}

\includegraphics{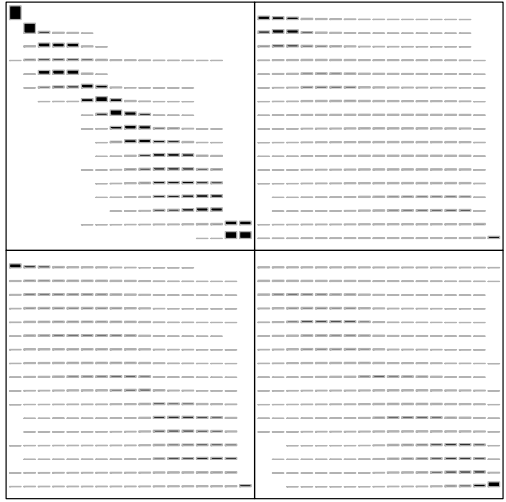}

\caption{Graphical testing of the $z$-matrix printed out in
Supplementary Table 4.
The top-left graphic shows the $z$-matrix (not transposed as in
Supplementary Table 4)
where each cell is represented by a rectangle with area proportional
to $z_{ij}$. The rows are therefore histograms with the same total area
for each row. The other three graphics are simulated $z$-matrices using
the overall $\hat{u}$, obtained by pooling all the data. The same
number of screens ($n_i$) were simulated for each reader $i$ as in the
data, and the matrices were ordered descending by simulated $\hat
{u}_i$, for consistency with Supplementary Table 4.
This graphical test suggests that there is some evidence to reject a
null hypothesis that all readers have the same recall-but-overruled
rate.} \label{figgraphtest}
\end{figure}

The difference between the centers is backed up by the overall
$z$-matrix (Supplementary Table 3): it has a block structure by center.
Separate $z$-matrices were produced to further investigate possible
differences between readers within each center.
The $z$-matrix for center 1 is in Table \ref{tblzijcat1}. It shows
that the readers with 0.5, 5~and 14 years experience appeared to be
different from the other readers. It can be seen from the $z$-matrix in
Table \ref{tblzijcat2} that readers in center 2 were harder to tell
apart, but there were possibly two distinct groups. However, these did
not appear to be correlated with reader experience. In passing, we note
that little attention should be paid to the reader $i$ with 0.5 years
experience because $z_{+i}-z_{ii}=0$ and so their parameter fit is
incompatible with all other readers' data; but $z_{ij} > 0$ for all $j$
with $z_{jj} = 0.489$ and so $i$'s data are not incompatible with the
other readers' parameters. This asymmetry occurs because they read
relatively few mammograms (Supplementary Table 2).
Table \ref{tblzijcat3}
shows the $z$-matrix for center 3. This is quite different to the
other centers because the concentration measures $z_{ii}$ are high for
all except one reader. Since the readers saw a similar number of
screens to the other centers, a systematic effect is likely to be
present within the center.
Further inspection of experience against the $\hat{\mathbf{u}}_i$'s
within center 3 showed a potential link between $S_{13}$ and reader
experience. To obtain further understanding, the categorical response
was dichotomized into $S_{13}$ against the rest, and a $z$-matrix for
centers 1 and 3 was obtained (all readers in center 2 had $S_{13}=0$),
as shown in Supplementary Table~4.
The ordering of individuals by their estimate $\hat{u}_i$ appears to
relate to experience shown in the second row of the table, and the
matrix pattern is inconsistent with a null hypothesis where everyone
has the same $u_i$ (Figure~\ref{figgraphtest}). The correlation to
experience is most clearly displayed in Figures~\ref{figonon}(b),~(c).

%
\begin{table}
\caption{($z^T \times1\mbox{,}000$)-matrix for dual-reader categorical
outcomes within center 2. The data are in Supplementary Table 2}
\label{tblzijcat2}
\begin{tabular*}{\tablewidth}{@{\extracolsep{\fill}}ld{3.0}d{3.0}d{3.0}
d{3.0}d{3.0}d{3.0}d{3.0}d{3.0}d{3.0}@{\hspace*{-2pt}}}
\hline
\multicolumn{1}{@{}l}{\textbf{Experience: }
\hspace*{2.4pt}\textbf{3}}&\multicolumn{1}{c}{\hspace*{-2pt}\textbf{7}}&
\multicolumn{1}{c}{\hspace*{-2pt}\textbf{22}}&\multicolumn{1}{c}{\hspace*{-2pt}\textbf{5}}
&\multicolumn{1}{c}{\hspace*{-2pt}\textbf{4}}&\multicolumn{1}{c}{\hspace*{-2pt}\textbf{4}}&
\multicolumn{1}{c}{\hspace*{-2pt}\textbf{6}}&\multicolumn{1}{c}{\hspace*{-2pt}\textbf{8}}&
\multicolumn{1}{c}{\hspace*{-2pt}\textbf{17}}&\multicolumn{1}{c@{}}{\textbf{0.5}}\\
\hline
\multicolumn{1}{@{}l}{\hphantom{\textbf{Experience: }}746}&214&159&6&&&&&&78\\
\multicolumn{1}{@{}l}{\hphantom{\textbf{Experience: }}176}&475&179&7&3&22&24&12&&62\\
\multicolumn{1}{@{}l}{\hphantom{\textbf{Experience: 0}}75}&170&446&107&9&8&2&3&&71\\
\multicolumn{1}{@{}l}{\hphantom{\textbf{Experience: 00}}3}&32&162&867&48&2&&1&&69\\
&3&20&12&835&35&4&26&2&55\\
&43&21&&72&437&188&207&77&55\\
&53&9&&12&192&425&225&123&44\\
&10&2&&20&242&263&309&298&43\\
&1&&&2&62&93&216&499&35\\
&&&&&&&&&489\\
\hline
\end{tabular*}
\end{table}

%
\begin{table}[b]
\caption{($z^T \times1\mbox{,}000$)-matrix for dual-reader categorical
outcomes within center 3. The data are in Supplementary Table 2}
\label{tblzijcat3}
%
\begin{tabular*}{\tablewidth}{@{\extracolsep{\fill}}ld{3.0}c
cccccc@{}}
\hline
\multicolumn{1}{@{}l}{\textbf{Experience: }
\hspace*{2.4pt}\textbf{0.1}}&\multicolumn{1}{c}{\hspace*{-2pt}\textbf{0.25}}
&\multicolumn{1}{c}{\textbf{2}}&\multicolumn{1}{c}{\textbf{3}}
&\multicolumn{1}{c}{\textbf{4}}&\multicolumn{1}{c}{\textbf{6}}
&\multicolumn{1}{c}{\textbf{9}}&\multicolumn{1}{c}{\textbf{10}}
&\multicolumn{1}{c@{}}{\textbf{18}}\\
\hline
\multicolumn{1}{@{}l}{\hphantom{\textbf{Experience: }}1\mbox{,}000}&2&&&&&&&\\
&770&&&&&&&\\
&53&980&\hphantom{0}18&&&&\hphantom{0}44&\\
&31&\hphantom{0}14&982&&&&&\\
&75&&&1\mbox{,}000&&\hphantom{0}11&&\\
&19&&&&994&\hphantom{0}65&\hphantom{0}16&\\
&19&&&&\hphantom{00}6&924&&\\
&31&\hphantom{00}5&&&&&940&\\
&&&&&&&&1\mbox{,}000\\
\hline
\end{tabular*}
\end{table}

%
%

\subsection{\texorpdfstring{Interpretation.}{Interpretation}}\label{secreader2Interp}

The readers in each center worked independently, and made their recall
decisions on their own. However, in center 2 the arbitration process
involved discussion between several readers, once a disagreement was
found between the first and second reader. This might be why readers in
center 2 were recorded as first or second reader after the outcomes had
been observed, and why the data (Supplementary Table 2)
show that when a reader in center 2 flagged a case for recall, the
case was always recalled regardless of the other reader. In any case,
it is clear that, as originally recorded, it is difficult to compare
readers from center 2 with the others in this analysis, and so it is
reasonable to leave them out of Figure \ref{figonon}.

The statistical structure shown by the $z$-matrix exploration and in
Figure \ref{figonon}, where the less-experienced readers tended to be
overruled more often, fits with a training effect. It is common
practice in dual reading to pair experienced readers with less
experienced ones. Thus, the increased rate of overruled recall flags
(by 3 different readers: the other, generally more experienced dual
reader, an arbitrator and the independent reader with CAD) might be
linked with less experienced readers being more cautious in their
recall decision. Overall, dual reading mitigates this by pairing
inexperienced readers with experienced ones who are able to overrule
unnecessary recalls. It is unclear whether single reading with CAD
would similarly mitigate this because, although the average experience
of readers using CAD in CADET II was similar to dual reading, the
minimum experience was 4 years (compare Figure \ref{figonon}). Thus,
in any implementation of screening based on a single reader with CAD,
it might be worth monitoring recall rates for readers with less than 4
years experience.

\section{\texorpdfstring{Dual reading vs. CAD false recall.}{Dual reading vs. CAD false
recall}}

So far we have considered analysis of the two screening regimens
separately. Further modeling may be used to look at them together. We
end by investigating the difference in recall rate between single
readers with CAD (3.9\%) and dual reading (3.4\%). To show the
technique from a different angle, we proceed as if we did not know
about the $z$-matrix, and first fit a statistical model to the data.
Then, the $z$-matrix will be used to help provide more understanding of
what the model has found.

\subsection{\texorpdfstring{Data.}{Data}}

Let $y_{ik} = 1$ if CAD reader $i$ recalls case $k=1,\ldots,n_i$ where
no cancer is detected on recall but dual reading does not, and $y_{ik}
= 0$ if CAD reader $i$ does not recall the case where no cancer is
detected on recall but dual reading does. Note that the comparison to
be made is between the cases where single readers with CAD or dual
readers flag for recall in error (but not both of them). The data from
%
%
\begin{table}
\tablewidth=260pt
\caption{Number of noncancers recalled by CAD reader ($y_{i+}$) when
dual readers did not recall, for all cases recalled in error by either
CAD readers or dual readers (but not both)} \label{tbllastanalysis}
\begin{tabular*}{\tablewidth}{@{\extracolsep{\fill}}lcccc@{}}
\hline
\textbf{Center} & \multicolumn{1}{c}{\textbf{Experience}}
& \multicolumn{1}{c}{$\bolds{y_{i+}}$} & \multicolumn{1}{c}{$\bolds{n_i}$} &
\multicolumn{1}{c@{}}{$\bolds{\hat{u}_i}$} \\
\hline
1 & \hphantom{0}4 & 21 & \hphantom{0}43 & 0.488 \\
1 & \hphantom{0}6 & 20 & \hphantom{0}59 & 0.339 \\
1 & 12 & 29 & \hphantom{0}50 & 0.580 \\
1 & 14 & 17 & \hphantom{0}32 & 0.531 \\
1 & 15 & 13 & \hphantom{0}28 & 0.464 \\
[4pt]
2 & \hphantom{0}4 & 38 & \hphantom{0}65 & 0.585 \\
2 & \hphantom{0}4 & 27 & \hphantom{0}42 & 0.643 \\
2 & \hphantom{0}5 & 29 & \hphantom{0}45 & 0.644 \\
2 & \hphantom{0}5 & 28 & \hphantom{0}44 & 0.636 \\
2 & \hphantom{0}6 & 18 & \hphantom{0}35 & 0.514 \\
2 & \hphantom{0}7 & 26 & \hphantom{0}43 & 0.605 \\
2 & \hphantom{0}8 & 29 & \hphantom{0}42 & 0.690 \\
2 & 17 & 34 & \hphantom{0}42 & 0.810 \\
2 & 22 & 38 & \hphantom{0}62 & 0.613 \\
[4pt]
3 & \hphantom{0}4 & 35 & \hphantom{0}92 & 0.380 \\
3 & \hphantom{0}6 & 46 & \hphantom{0}96 & 0.479 \\
3 & \hphantom{0}9 & 61 & 103 & 0.592 \\
3 & 18 & 45 & \hphantom{0}88 & 0.511 \\
\hline
\end{tabular*}
\end{table}
the trial are shown in Table \ref{tbllastanalysis}: if $\hat{u}_i <
0.5$, then the CAD reader did better than dual readers, and if $\hat
{u}_i > 0.5$, then they did worse.

\subsection{\texorpdfstring{Model.}{Model}}

Consider a model
\[
\operatorname{logit}\{P(y_{ik}=1|\mathbf{x}_{i},v_i;\bolds{\beta
},\sigma^2)\} = \mathbf{x}_{ik}\bolds{\beta}' + v_i,
\]
where $\bolds{\beta} = (\beta_0,\beta_1,\beta_2)$ are parameters
and $\mathbf{x}_{i} = (x_{i1},x_{i2},x_{i3})$ are covariates; $v_i$
is a random effect taken (for convenience) to be from a Normal
distribution with mean 0 and variance $\sigma^2$; and $
\operatorname{logit}(\cdot)$ denotes the logistic function.
The covariates are a constant ($x_{i1}=1$), a factor for center 2
($x_{i2}=1$ for center 2, 0 otherwise) and a factor for reader
experience ($x_{i3})$, whose form is explored below. Thus, the baseline
is for centers 1 and 3 and readers with the reference reader
experience. Other covariates [about the screen: first ever screen
(incident) or not (prevalent), age, a score from the CAD algorithm
predicting the likelihood of cancer; and about the reader: training
(radiographer, radiologist, other)] were explored but did not
significantly improve the model fit.

Maximum-likelihood estimation (the routine \texttt{xtlogit} in the
computer software \texttt{STATA} that uses Gauss--Hermite quadrature for
the likelihood) is used to find odds ratios and Wald 95\% confidence
bounds on the effects. The first definition of reader experience is a
binary variable $x_{i3}=1$ when reader~$i$ has more than six years
experience, 0 otherwise. This definition was chosen because it roughly
balances the readers by center, as seen in Table \ref
{tbllastanalysis}. The estimated odds ratios for center 2 and reader
experience effects are, respectively, $_{{1.54}}{2.01}_{{2.61}}$ and
$_{{1.23}}{1.59}_{{2.06}}$, where we use the useful notation from
\citet{2009LouisZeger} to present the point estimate surrounded
by a
95\% confidence interval. Using this definition of experience seems to
account for most between-reader variation because $\ln(\hat
{\sigma}^2) = {-13.6}_{(43.0)}$ [again following \citet
{2009LouisZeger} to put the standard error as a subscript]. Indeed,
identical odds ratios are found from a straight logistic regression
without~$v_i$.
Other definitions of reader experience suggest that a linear
relationship is not a good one: if years of experience are used, then
the odds ratio estimate is $_{{1.00}}{1.02}_{{1.05}}$, and the
random-effect term becomes more important with $\ln(\hat
{\sigma
}^2) = {0.15}_{(0.01)}$. Another possibility is to use $
\log(\mathrm{experience})$, which resulted in an estimated reader
experience odds ratio of $_{{1.04}}{1.33}_{{1.70}}$.

The model fits provide some evidence that, perhaps surprisingly, the
less experienced readers were less likely to recall in error with CAD
than the experienced ones. This is different to the trend seen in
\citet
{2006Astleyetal}, although that was a retrospective study. Taken
together with the results in Section \ref{secreader2Interp}, this
might be interesting because it suggests that CAD might help the less
experienced readers ($<$7 years) avoid unnecessary recall
decisions. However, given that $n=18$, one might be interested in
understanding more about the data's structure, especially given the
change in effect size depending on reader experience definition. We
will proceed to further investigate using the $z$-matrix and some of
the plots previously used.

\subsection{\texorpdfstring{$z$-matrix analysis.}{$z$-matrix analysis}}

\begin{figure}
\begin{tabular}{@{}cc@{}}

\includegraphics{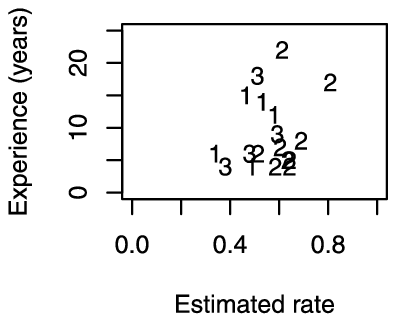}
 & \includegraphics{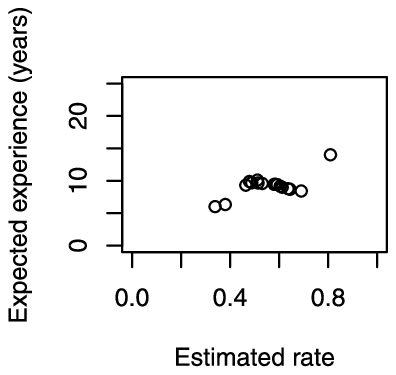} \\
(a) & (b)\\[4pt]

\includegraphics{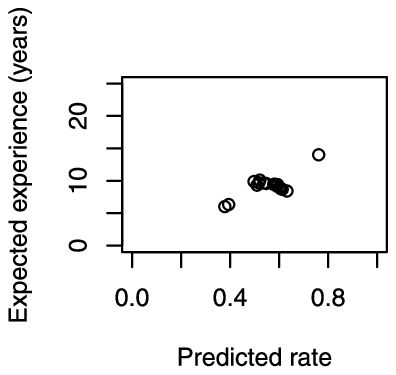}
 & \includegraphics{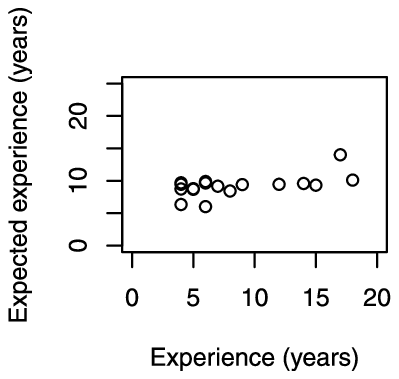} \\
(c) & (d)
\end{tabular}
\caption{Exploratory plots for single reader with CAD recall-in-error
rate (relative to double reading) vs. reader experience
(years). Plot \textup{(a)} shows experience against the original estimates $\hat
{u}$, with center number as the symbol. Plot \textup{(b)} uses expected
experience on the $y$-axis, following the approach in Section \protect
\ref
{secexasmoothx}. Plot \textup{(c)} replaces the original estimate $\hat{u}$ by
a~prediction from equation (\protect\ref{eqnprede}). Plot \textup{(d)}~shows how
expected experience relates to the original data.} \label{figonon2}
\end{figure}

The $z$-matrix is shown in Supplementary Table~5.
The data driving the experience effect from the model are that two
readers with 6 and 4 years experience have relatively low $\hat{u}_i$,
with their $z_{ij}$ close to the other's $z_{ji}$, and they are
relatively concentrated; and one reader with 17 years experience has
the highest $\hat{u}_i$, which is also more concentrated than those in
between. A center structure can also be observed: center 2 against the
others. The experience pattern is also seen in Figure \ref{figonon2},
%
where the three readers are clear in plots (b) and (c). However, some
caution in interpreting the reader experience correlation is required:
differences are seen between the null and observed $z$-matrices in
Supplementary Figure~2, but the pattern of two low $\hat{u}_i$ and a
single high $\hat{u}_i$ might be due to chance. The plot casts doubt on
whether the pattern is real, or whether it was (mis)fortune that led to
the reader experience effect. A $z$-matrix examination therefore showed
that the correlation between reader experience and $u_i$ was driven by
$3/18$ readers with behaviors in opposite directions, but also showed
that it is quite a weak finding.

\subsection{\texorpdfstring{Other techniques.}{Other techniques}}

The model fit may be explored in other ways. We end by using prediction
to show that the center 2 effect is a more robust finding.
If a noncancer case is not recalled by the single reader with CAD,
then, using centers 1 and 3, we fit a logistic-regression model for the
probability of recall by dual reading with covariates for incidence/prevalence (first or subsequent screen) and whether the case was
arbitrated. A prediction from this model is that 156 such cases could
be expected at center 2. This compares against an observed number of
130, so the dual readers did slightly better than might be expected. A
similar logistic-regression model was fitted to centers 1 and 3 for
recall by the single reader with CAD, given the case was not a cancer
and was not recalled by dual reading. A covariate for incidence/prevalence
status was used together with a continuous variable
correlated to the probability of cancer according to the computer tool.
This model predicted a total of 140 such cases, but 267 were observed.

\section{\texorpdfstring{Application of the exploratory approach to other data.}{Application of the exploratory approach to other
data}}

The $z$-matrix applies quite generally to the two-stage statistical
setup described in Section~\ref{secmethod}. A~similar data structure
is found in other applications, such as the effect of physical tasks of
patients, blood glucose levels and rat body weights that are in
\citet
{1990CrowderHand}; as well as many others including sport where
individuals have repeated attempts to, for example, hit a ball in
cricket, or score a goal in football; or in the workforce when
productivity is measured by number of items processed by the worker.
Thus, the technique might be used for growth curves, point processes or
any other data structure where it is possible to write down a
likelihood function for the individual.

One aim is to use the data to find structure among the units that would
be seen again in future samples. A common approach to this problem is
to fit a two-stage model, which, as seen in the above data analysis,
might produce similar findings to the $z$-matrix approach. However,
some strengths of the $z$-matrix as an exploratory technique, relative
to use of full statistical models, include the following:
\begin{itemize}
\item As seen in Section \ref{secmthdsim}, $z_{ij}$ is a comparative
measure that has a direct interpretation in terms of how close
individual $j$'s parameter fit is to individual~$i$'s data. Although
other approaches can be used to estimate $p(\mathbf
{u}_i|\mathbf
{y})$, they lose the direct comparative aspect that arises in the
$z$-approach from restricting the $\mathbf{u}$ support to only
contain $\hat{\mathbf{U}}_n = (\hat{\mathbf{u}}_1,\ldots
,\hat
{\mathbf{u}}_n)$. For example, when using NPML in Section~\ref
{secanal1} an equivalent ``$z_{ij}$'' would have $j=1,2$ because there
are two support points.
\item Plots such as Figure \ref{figxtraplot1} show that the $z$-matrix
can be used to provide an indication of how many distinct groups there
might be; NPML simply gives the most likely number. For exploratory
analysis both are useful.
\item The measure can be interpreted in a similar manner for different
$p(y|\mathbf{u})$ likelihoods, and the information from
$\mathbf
{u}_i$ vectors is shown in the same two-dimensional way for any
dimension of $\mathbf{u}_i$. That is, the approach standardizes
comparisons between the $\mathbf{u}_i$ vectors for different types
of response variables.
\item The approach is quite general and can be easily applied to
different $p(y|\mathbf{u})$ likelihoods. Although with a binomial
likelihood many other approaches are feasible using statistical
software, this will not always be the case. For example, the
$z$-approach was used for prediction when the likelihood function was
of a self-exciting point process form in \citet{2008BrentnallCrowderHand}.
\item For prediction the approach provides a simple approximate route
to BLUP's (best linear unbiased predictors), or posterior means,
through equation~(\ref{eqnprede}). Some evidence of the benefit of
predictions formed in this way using real data, compared with
parametric empirical Bayes predictions, is found in \citet
{2010BrentnallCrowderHand}.
\item Finally, while computationally-intensive methods may be justified
for statistical modeling, it seems much less attractive to have to wait
for exploratory analysis to run. Once the point estimates have been
obtained, the method requires $O(n^2)$ computations for equation (\ref
{eqnzij}). This makes it most appealing for small to moderate $n$.
\end{itemize}

\section{\texorpdfstring{Conclusion.}{Conclusion}}

In this work we developed a method of exploratory analysis for
applications in which repeated measurements have been recorded on
a~group of individuals. The aim of the approach is to draw attention to
groups of similar behaviors, outliers and trends in the data. It does
so by helping to quantify prediction uncertainty between individual
point estimates through a ``similarity'' measure. This $z$-matrix used
can be viewed as a discrete approximation to an empirical Bayes
posterior distribution.
The approach was motivated by an analysis of reader performance in
CADET~II. We showed its application to binary and multinomial response
variables, and illustrated some identified properties of the measure
using the data. One avenue for future research is to extend the
approach to explicitly account for more than two levels in the
hierarchical data structure. Such an extension would be useful for
cancer screening since readers are sampled from screening centers.

\section*{\texorpdfstring{Acknowledgments.}{Acknowledgments}}
We thank Professor Karen Kafadar and two anonymous referees for their
suggestions that have helped to improve the methodology and
presentation of the application in this article.

\begin{supplement}
\stitle{Supplement to ``A method for exploratory
repeated-measures analysis applied to a breast-cancer screening study''}
\slink[doi]{10.1214/11-AOAS481SUPP}
\slink[url]{http://lib.stat.cmu.edu/aoas/481/supplement.zip}
\sdatatype{.zip}
\sdescription{Some additional tables and charts to accompany this paper.}
\end{supplement}



%
\printaddresses

\end{document}